# Deformation and adiabatic heating of single crystalline and nanocrystalline Ni micropillars at high strain rates

**Nidhin George Mathews[1,a], Matti Lindroos[2], Johann Michler[3], Gaurav Mohanty[1]**

[1] *Materials Science and Environmental Engineering, Tampere University, Korkeakoulunkatu 6, Tampere 33014, Finland.*

[2] *VTT Technical Research Centre of Finland Ltd, Tekniikantie 21, Espoo 02044, Finland.*

[3] *Laboratory of Mechanics of Materials and Nanostructures, Empa, Swiss Federal Laboratories for Materials Science and Technology, Feuerwerkerstrasse 39, Thun 3602, Switzerland.*

[a] *Corresponding author: nidhin.mathews@tuni.fi*

## Abstract

The deformation behavior of single crystal and nanocrystalline nickel were studied using *in situ* micropillar compression experiments from quasi-static to high strain rates up to $10^3$ s$^{-1}$. Deformation occurred by dislocation slip activity in single crystal nickel whereas extensive grain boundary sliding was observed in nanocrystalline nickel, with a shift towards more inhomogeneous, localized deformation above 1 s$^{-1}$. The strain rate sensitivity exponent was found to change at higher strain rates for both single crystal and nanocrystalline nickel, while the overall strain rate sensitivity was observed to be of the same value for both. With increasing high strain rate micropillar compression tests being reported, the issue of adiabatic heating in micropillars becomes important. We report crystal plasticity based finite element modeling to estimate the adiabatic heating, spatially resolved within the pillar, at the highest tested strain rates. The simulations predicted a significant temperature rise of up to 200 K in nanocrystalline Ni at the grain boundaries, and 20 K in single crystalline Ni due to strain localization. Transmission Kikuchi Diffraction analysis of nanocrystalline nickel pillar post compression at $10^3$ s$^{-1}$ did not show any grain growth.

## 1 Introduction

The mechanical response of materials at high strain rates is a matter of great concern and of immense scientific interest for various engineering applications where impact loading conditions prevail, for e.g. in automobile crash testing and aerospace applications. Macroscale uniaxial high strain rate (HSR) testing is generally performed with ballistic devices such as split-Hopkinson/Kolsky bar setups with speeds up to 100 ms$^{-1}$ and strain rates up to 10,000 s$^{-1}$ [1,2]. Moderately high strain rates can be achieved with servo-hydraulic testers reaching speeds of 10 ms$^{-1}$. Charpy and Izod impact testers are used for determining the fracture behavior and impact resistance at moderately high strain rates. However, all these tests provide an averaged microstructural response which does not allow investigating the influence of grain anisotropy and grain boundaries individually. While applying digital image correlation on speckle patterns allows the study of local evolution in strains with high lateral resolutions inside the grains and along grain boundaries at quasistatic strain rates, however, it is not straightforward to apply this approach to split-Hopkinson pressure bar (SHPB) tests due ultra-high speeds of testing. Investigating the response of individual phases, grain orientations and grain boundaries of bulk materials is necessary for developing reliable multi-scale models to optimize phase distribution and microstructure for impact applications. Also, with increasing degree of miniaturization, it becomes necessary to perform small scale, high strain rate tests to assess the impact resistance and crash worthiness of thin films and micro-electro-mechanical systems (MEMS) to design durable devices.

Nanoindentation is the most widely used small scale mechanical testing technique due to minimal sample preparation, offering the possibility to perform hundreds of repeats on the same sample and offering high lateral, depth and temporal resolution mechanics data. However, constant strain rate nanoindentation has traditionally been confined to quasi-static strain rates (QSR) (typically from $10^{-4}$ s$^{-1}$ to 1 s$^{-1}$). Although nanoindentation impact tests have existed for close to three decades, these tests are not constant strain rate tests, like their macroscale counterparts, and suffer from rapidly decreasing indentation strain rate with depth [3,4]. This makes the assignment of an effective strain rate and the determination of the corresponding hardness value somewhat unreliable. Also, pyramidal indentation, with large stress gradients, can potentially result in fundamentally different deformation mechanisms than those expected under uniaxial stresses. For e.g. Zhang et al. reported extensive grain boundary formation in aluminum but not so much in molybdenum from high strain rate Berkovich nanoindentation tests [5]. Micropillar compression tests performed with a nanoindenter using a flat punch tip geometry allow probing uniaxial response at small length scales. Recent developments in nanoindentation hardware and control software capabilities have pushed the envelope of high strain rate testing up to ~ $10^3$ s$^{-1}$ [6–9] by overcoming resonance issues at high speeds. For e.g. Cherukuri et al. reported constant strain rate nanoindentation up to $10^3$ s$^{-1}$ and micropillar compression up to ~ 750 s$^{-1}$ [8]. This opens up the exciting possibility for exploring the high strain rate deformation behavior at small length scales. There is still a clear lack in our understanding of possible changes in deformation mechanisms at high strain rates in this rapidly emerging, but nascent, field of high strain rate micromechanics research. The combination of small volumes in case of micropillar compression, which allows dislocation exhaustion at the pillar surface, and high strain rates can potentially result in changes in strain rate sensitivity and deformation mechanism transitions at high strain rates. For this study, we have chosen a model fcc metal – nickel – that is available in both single crystalline (sx) and nanocrystalline (nc) form, to systematically investigate uniaxial micropillar compression covering strain rates from $10^{-3}$ s$^{-1}$ to ~ $10^3$ s$^{-1}$. We use *in situ* testing inside a scanning electron microscope (SEM) that allows us to precisely compress focus ion beam (FIB) milled micropillars to ~ 20% strain and observe their deformation behavior. We pay specific attention to possible changes in strain rate sensitivity exponent and secondary electron (SE) imaging of the deformed pillar surface to probe deformation mechanism changes, if any.

A secondary objective of this study is to estimate the extent of adiabatic heating in these micropillars using crystal plasticity based finite element (CP-FE) modeling. It is well known at macro-length scales that a fraction of the input mechanical work is converted to heat during high strain rate testing. Due to the short duration of such tests, the heat produced during the process is not dissipated fast enough to the surroundings. Hence, it is considered to be adiabatic in nature. The extent of adiabatic heating in high strain rate tests is typically studied using the Taylor-Quinney coefficient ($\beta$), which provides the fraction of plastic work that gets converted to heat during rapid deformation of the material [10]. The Taylor-Quinney coefficient has been shown to reach values of ~ 0.9 for strain rates above $10^3$ s$^{-1}$ and at high strains with temperature increase above 50 ºC reported in some FCC metals [11–13]. Specifically, temperature increase of 5 - 70 ºC was reported for FCC copper at strain rates above $10^3$ s$^{-1}$ [11,14]. Soares et al. reported a Taylor-Quiney coefficient change from 0.6 to 0.9 in FCC copper for strain rates from 1300 s$^{-1}$ to 3100 s$^{-1}$ [14]. Large adiabatic heating can potentially influence thermally activated deformation processes, such as dislocation nucleation, glide and cross-slip, and strain hardening [15]. It can also result in microstructural changes in the deforming volume for e.g. grain growth and new boundary formation. Therefore, it is necessary to understand the extent of adiabatic heating in micropillar compression at high strain rates and its potential impact on mechanical response. High-speed thermal cameras are typically used to experimentally determine adiabatic heating related temperature rise in macroscale high strain rate tests [16]. However, such thermal cameras are not suitable for micropillar compression where the tested volumes typically vary from a few to tens of μm$^3$. For context, state-of-the-art infrared cameras offer lateral resolutions of ~ 5-20 μm only [17–19], which is larger than the pillar diameters used in majority of microcompression studies. Also, the shielding by the indenter tip, which typically hovers couple of 10's of nm above the pillar surface before the compression tests, and the surrounding material, as the pillars are milled using a FIB by creating a crater around it, make it additionally difficult to focus the thermal cameras onto the pillar side wall during the test. Therefore, existing literature on microscale testing typically ignores the issue of adiabatic heating and assumes no influence on mechanical deformation behavior. Ramachandramoorthy *et al*. estimated temperature rise of 110 ºC in fused silica glass micropillars at $10^3$ s$^{-1}$ using simple thermal finite element simulations [7]. No such estimates on adiabatic heating associated temperature rise exist for microscale compression of metals, to the best of our knowledge. To fill this critical gap, we utilize a comprehensive CP-FE model to estimate temperature rise in both sx and nc Ni pillars at the highest tested strain rate, with high spatial resolution within the pillars.

In this study, we report micropillar compression data on single crystal and nanocrystalline nickel covering seven orders of strain rates from $10^{-3}$ s$^{-1}$ to $10^3$ s$^{-1}$. The choice of Ni is motivated by face centered cubic (FCC) crystal structure, ease of availability of a microstructurally stable nanocrystalline form through electrodeposition and relatively high yield strength of its nanocrystalline form which promotes substantial adiabatic heating. The objectives of this study are two fold: (a) to understand the deformation behavior and possible deformation mechanism changes in micropillar compression of sx and nc Ni, across tested strain rates, and (b) to compare the extent of adiabatic heating between sx and nc Ni micropillars at the highest tested strain rates. The strain rate sensitivity exponents, apparent activation volumes and surface features on the micropillars post-deformation are analyzed across the tested strain rates. The extent of temperature rise in the pillars due to adiabatic heating in these two materials is estimated using a size-dependent micromorphic CP-FE model coupled with adiabatic heating which converts the non-stored mechanical energy to heat. This is the first application of such a comprehensive CP-FE model to micropillar compression on metals which will allow us to understand the temperature hotspots within the microstructure during high strain rate deformation. Finally, transmission Kikuchi diffraction (TKD) is performed on lift-outs from a $10^3$ s$^{-1}$ compressed nc Ni pillar to study any possible grain growth due to adiabatic heating.

## 2 Materials and methods

### 2.1 Specimen details

Single crystal nickel (sx Ni) sample of (100) plane orientation was purchased from the commercial supplier (MTI Corporation, USA). Nanocrystalline nickel (nc Ni) was manufactured using a proprietary electrodeposition technique on silicon wafer. The structural characterization details of the nc Ni sample are reported in a previous work [20]. The silicon wafer was later dissolved in hydrofluoric acid to obtain free standing nc Ni section. The thickness of the free-standing layer was measured to be 0.25 mm. A standard FIB-lamella lift out procedure was performed on the nc Ni sample using the FIB-SEM (Zeiss Crossbeam 540, Germany) to prepare the sample for the transmission electron microscope (TEM) characterization (Jeol JEM-F200, Japan) and transmission Kikuchi diffraction (TKD). TEM micrographs and electron diffraction patterns displayed the nanocrystalline grain structure (*supplementary information S1*). The inverse pole figure (IPF) map determined from TKD (Zeiss Ultraplus, Germany) of the nc Ni sample showed the random orientation of grains with an average grain size ($\delta$) of 23 ± 3 nm.

### 2.2 Micropillar milling and micromechanical testing

Micropillars of ~ 3 μm diameter (*d*) with aspect ratio (*l/d*) between 3 - 3.5 were fabricated using focused ion beam (FIB) milling (Zeiss Crossbeam 540, Germany) using Ga$^+$ ions. Ion currents of 7 nA, 3 nA and 300 pA at 30 kV accelerating voltage were used for coarse, intermediate, and fine milling steps respectively in order to progressively reduce Ga$^+$ implantation damage. Secondary electron (SE) images of micropillars were captured after milling, from which a taper angle less than 2º was measured for all the micropillars. The *in situ* compression of these micropillar samples were performed inside the scanning electron microscope (Zeiss Leo 1450, Germany) using the displacement-controlled indenter (Alemnis AG, Switzerland). Micropillars were compressed uniaxially with a diamond flat punch indenter tip (Synton MDP, Switzerland) of diameter 5 μm up to 20% nominal strain at orders of strain rates ranging from $10^{-3}$ s$^{-1}$ to $10^3$ s$^{-1}$. Standard load cell assembly was used for micropillar compression in the quasi-static strain rate regime ($10^{-3}$ s$^{-1}$ to 10 s$^{-1}$), whereas a piezo-based smart-tip load cell assembly was used for HSR regimes (3 s$^{-1}$ to $10^3$ s$^{-1}$) [6]. The cross-sectional area at the top of the pillar was considered to determine the stresses from the recorded load data, with the flow stress at 1% strain offset is expressed as the yield strength of the material for all further analysis. The consistency of the yield strength value measured from the standard assembly and the smart-tip assembly was cross-checked at overlapping strain rates of 3 s$^{-1}$ and 10 s$^{-1}$ of both setups. The average yield strength value obtained from standard setup was found to be ~ 6% lower than the smart-tip setup. Therefore, all yield strength values measured from the smart-tip setup were normalized to the standard setup, in order to be consistent with the measurements. The strain rate sensitivity ($m$) and apparent activation volume ($V_{app}$) for deformation were calculated using Eq. 1 and Eq. 2 respectively using 1% offset yield stress data.

$$m = \left[\frac{d(\ln \sigma)}{d(\ln \dot{\varepsilon})}\right]_T \qquad Eq.\ 1$$

$$V_{app} = \frac{\sqrt{3}kT}{m\,\sigma} \qquad Eq.\ 2$$

where, $\sigma$ is the flow stress, $\dot{\varepsilon}$ is the strain rate, $k$ is the Boltzmann constant, $T$ is the absolute temperature.

## 2.3 Crystal plasticity modelling

Size-dependent crystal plasticity model was employed with a micromorphic strain gradient extension to investigate low and high strain rate deformation behavior of nickel. Finite strain formalism was used with decomposition of deformation gradient into an elastic part, $\underline{F}^E$ and a plastic part, $\underline{F}^P$, as shown in Eq 3.

$$\underline{F} = \underline{F}^E \cdot \underline{F}^P. \qquad Eq.\ 3$$

A dislocation density-based formulation is utilized which shares similarities with an FCC model suggested by Monnet et al. [21]. Plastic velocity gradient $\left(\underline{L}^p\right)$ is computed as the sum of all slip systems, as shown in Eq. 4.

$$\underline{L}^p = \sum_{s=1}^{N_s} \dot{\gamma}^s \, \underline{N}^s \qquad Eq.\ 4$$

where, $\underline{N}^s$ is the orientation tensor of a slip system $s$. Plastic deformation in FCC Ni is carried over twelve octahedral $\{1\,1\,1\}\langle 1\,1\,0\rangle$ slip systems in the model. Slip rate of a dislocation slip system $s$ is given by a Norton-type of flow rule in Eq. 5.

$$\dot{\gamma}^s = \left\langle \frac{|\tau^s| - \tau_c}{K} \right\rangle^N sign(\tau^s) \qquad Eq.5$$

where $\tau^s$ is the resolved shear stress of a slip system computed with $\tau^s = \left(\underline{C}^E \cdot \left(\underline{\underline{A}} : \underline{E}_{GL}\right)\right) : \underline{N}^s$, where $\underline{C}^E$ is the Cauchy-Green tensor, $\underline{\underline{A}}$ is the stiffness tensor, and $\underline{E}_{GL}$ is the Green-Lagrange strain tensor. Parameters $K$ and $N$ describe viscosity and strain rate dependency, respectively. Slip resistance of a slip system is defined in Eq. 6 as:

$$\tau_c^s = \tau_0 + \tau_{HP} + \mu b^s \sqrt{\sum_{s=1}^{N_s} a_{eff}^s \rho^s} - S_\chi \qquad Eq.6$$

where, $\tau_{HP} = \frac{\mu}{\mu_{RT}} \frac{K}{\sqrt{d}}$ is the Hall-Petch effect. Coefficient $K$ is the Hall-Petch constant and effective grain size is denoted with $d$. The Hall-Petch effect in the model is scaled with a temperature-dependent shear modulus, $\mu$, with respect to room temperature shear modulus $\mu_{RT}$.

Solid solution strengthening term $\tau_0$ is assumed to be zero for the pure metal and the slip resistance of the material is dependent on the initial dislocation density and the related interactions between slip systems. Generalized stress is included in the model using reduced micromorphic formulation, as described in detail in [22–24]. The model introduces size effects on the material model and regularizes slip localization. The generalized stress is defined in Eq. 7 as:

$$S_\chi = -H_\chi(\gamma_{cum} - \gamma_\chi) = A\Delta_\chi \gamma_\chi \qquad Eq.7$$

where two length-scale parameters are introduced $H_\chi$ and $A$. Cumulative plastic slip is computed by $\gamma_{cum} = \int_0^t \sum_{s=1}^{N_s} |\dot{\gamma}^s|\, dt$.

The dislocation interactions coefficients are not constant in the model, and they evolve similarly as suggested in Eq. 8 [21].

$$a^s_{eff} = \left(0.2 + 0.8 \frac{\ln(b^s \sqrt{\sum_{s=1}^{N_s} a^s_{const} \rho^s})}{\ln(b^s \sqrt{\rho_{ref}})}\right)^2 a^s_{const} \qquad Eq.8$$

Dislocation density evolution is described in Eq. 9:

$$\dot{\rho}^s = \frac{|\dot{\gamma}^s|}{b^s}\left[\frac{1}{d} + \frac{\sqrt{\sum_{forest} a^s_{eff}\rho^s}}{K_{forest}} + \frac{\sqrt{\sum_{coplan} a^s_{eff}\rho^s}}{K_{coplan}} - y\rho^s\right]d \qquad Eq.9$$

where hardening related parameters $K_{forest}$ and $K_{coplan}$ describe the number of obstacles overcome by dislocations on average before becoming immobile. Dislocation annihilation is formulated with temperature and strain rate dependent distance $y$ [25], defined in Eq. 10.

$$y = 2 y_0 \left(1 - \frac{k_B T}{A_{rec}} \ln\left|\frac{\dot{\gamma}^s}{\dot{\gamma}_0}\right|\right) \qquad Eq.10$$

where $y_0$ is reference annihilation distance, $A_{rec}$ is the capture radius energy, and $\dot{\gamma}_0$ is the reference slip rate. Adiabatic heat generation and the related increase rate of temperature is similar to [24,26], but modified for the micromorphic model as given in Eq. 11.

$$\dot{T} = \frac{\beta(\sum_{s=1}^{N_s} \tau^s \dot{\gamma}^s + S_\chi \gamma_{cum})}{\rho c} \qquad Eq.11$$

where $\beta$ is the Taylor-Quinney coefficient (assumed constant 0.9 here), $\rho$ is the mass density, and $c$ material's heat capacity. It is assumed that for strain rates > $10^2$ s$^{-1}$, the condition is fully adiabatic. Thermal strains will build up in the material due to uneven heating, however, their effect is omitted in the current study for simplicity.

A sensitivity analysis was performed on adiabatic heating of micropillar and a nano-grained polycrystalline material. Table S1 (supplementary information S2) lists the used crystal plasticity parameters. The model parameters were identified using low and high strain rate micropillar compression experimental data at $10^{-2}$ s$^{-1}$ and $10^2$ s$^{-1}$, respectively. Simulations were performed up to 15% of strain and the orientation of the compressive loading was [100] for sx Ni. To investigate the adiabatic heating effects in nano-grained polycrystalline nickel, a polycrystalline simulation case with 200 grains was used. Compression was applied similarly up to around 15% of axial strain, as in the single crystal. The grains were randomly oriented, and the grain size was set to 30 nm, as reported in a previous study on the same batch of electrodeposited nc Ni sample [20]. An initial dislocation density of $4.8 \times 10^{15}$ m$^{-2}$ was assigned for the polycrystal. Simulations on a micropillar fully populated with 30 nm grains is computationally very expensive to investigate, and therefore we used a simpler 3D domain for the analysis.

## 3 Results and Discussion

### 3.1 Effect of strain rates on the yield strength

The engineering stress-strain response of the sx and nc Ni samples determined from the micropillar compression experiments at various orders of strain rates are shown in Fig. 1a and Fig. 1b. Both sx and nc Ni showed plastic deformation after the linear elastic behavior. The sx Ni showed typical serrated response in the plastic region which is the characteristic deformation behavior due to dislocation avalanches. Yield strength values of sx Ni were observed to range between 100 MPa to 200 MPa across the tested strain rates. The serrated plastic flow due to the dislocation avalanches is more pronounced at strain rates from $10^{-3}$ s$^{-1}$ to $10^{-1}$ s$^{-1}$, whereas at strain rates above 1 s$^{-1}$, the serrated response gets camouflaged within the load noise level. It should be noted that the micropillar compression experiments on sx Ni are reported only up to $10^2$ s$^{-1}$ strain rate due to low signal-to-noise-ratio observed at higher strain rates (i.e. at $10^3$ s$^{-1}$). The tested pillars had a small variation in pillar diameter of <150 nm which can potentially affect the observed yield strength. Based on the works of Uchic et al. [27,28] which have reported a systematic dependance of yield strength on the pillar diameter for sx Ni and size scale exponent of ~ 0.6 - 0.7, this pillar diameter variation of ~ 150 nm was found to cause a yield stress difference of < 5%. Sx Ni pillars show increased hardening behavior with strain rate. This suggests increased dislocation-dislocation interactions and potentially dislocation substructure formation in our ~ 3µm diameter sx Ni pillars. Zhao et al. have reported deformation induced dislocation substructure formation in copper micropillars with diameters of 3 µm and above [29]. It is well known that dislocation starvation occurs for pillar diameters of 1 µm and less whereby the dislocations escape to the pillar free surface. Zhao et. al demonstrated that for larger diameters, sufficient volume is available for substantial dislocation-dislocation interactions to occur resulting in dislocation cell formation. They also observed slightly increased dislocation cell formation from $10^{-4}$ to $10^{-2}$ s$^{-1}$ [29]. Zhang et al. reported similar trends in increased dislocation substructure formation in aluminium using Berkovich nanoindentation [5].

The dislocations can freely escape to the pillar surface in sx Ni micropillars, whereas in nc Ni, the grain boundaries restrict the flow of the dislocations to the free surface. The hardening effect in sx Ni is also observed due to the creation of geometrically necessary dislocations to accommodate the imposed strain which was observed. This increases the hardening rate with decreasing pillar sizes and is more evident in sub-micrometer pillars [30,31]. The yield strength of nc Ni is ~ 15 times that of sx Ni due to Hall-Petch strengthening, which are in good agreement with previous studies [20]. In contrast to sx Ni, the hardening behavior of nc Ni seems to be fairly constant for all strain rates [32]. This is due to the inability of the nanocrystalline grains to accommodate multiple dislocations. That's why nanocrystalline FCC metals do not typically exhibit strain hardening and serrated behavior. The grain boundary mediated plasticity in nc metals consists of constant generation and annihilation of dislocations at the grain boundaries. It should be noted that the small hardening observed in the plastic regime of nc Ni pillars should be considered as an apparent hardening effect which arises due to geometric taper of the FIB milled micropillars [33].

Strain rate dependance of sx and nc Ni at microscale has been previously studied by strain rate jump tests [20,33] and constant strain rate tests [6,34,35] with indentation and micropillar compression experiments. These varying strain rate tests provided insights to the underlying deformation mechanisms and thermal activation during deformation from the strain rate sensitivity and activation volumes estimated. However, in microscale all of these tests were limited to quasistatic strain rate regime only and the possibility of high strain rate of deformation were not explored.

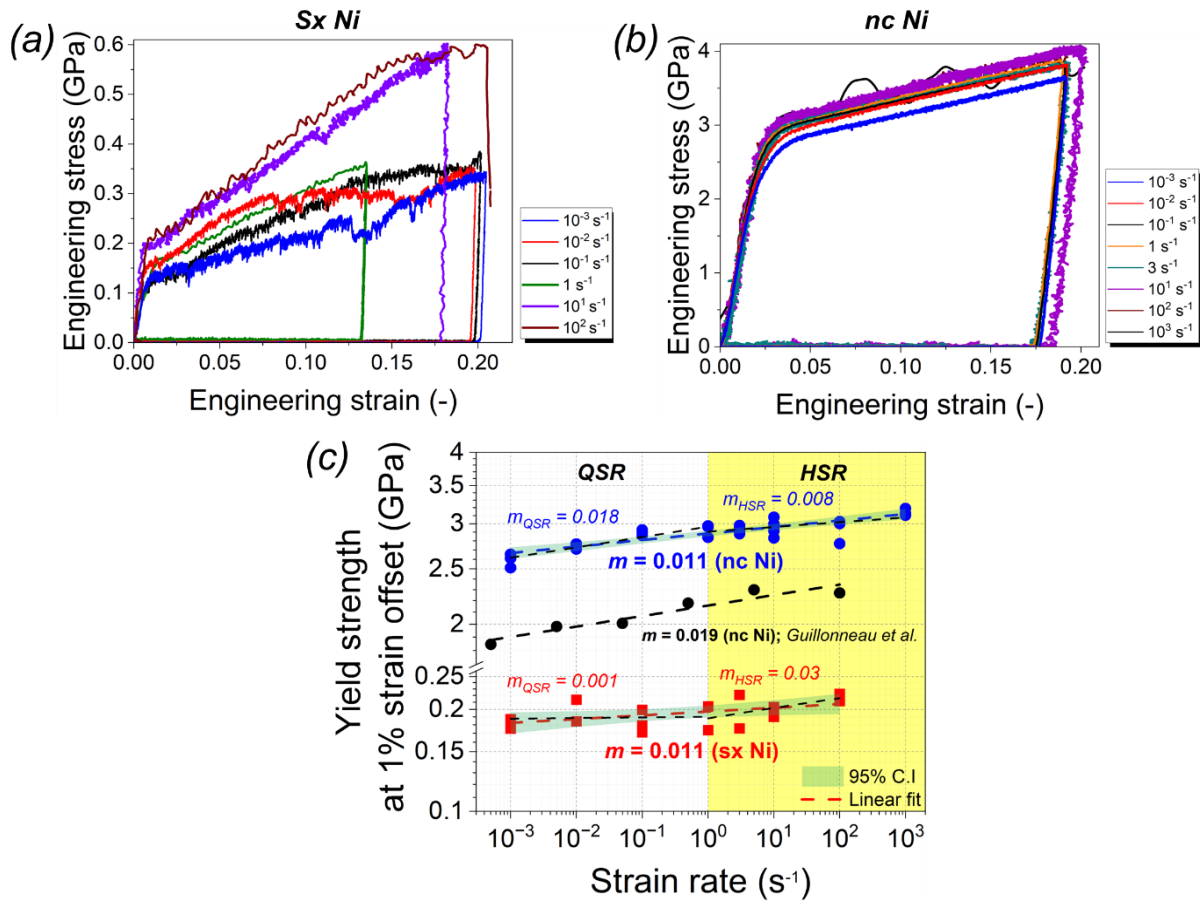

*Fig. 1: Engineering stress-strain curves obtained from micropillar compression at different strain rates for a) sx Ni and b) nc Ni. c) Yield strength values of sx and nc Ni at quasistatic and HSR strain rate regimes with corresponding strain rate sensitivity values. Results from a similar previous work [6] is also replotted for comparison. (Note: Dashed line represents the linear fit of the data points).*

The logarithmic plot of yield strength values at 1% strain offset at the corresponding strain rates of deformation is shown Fig. 1c. These *m* values obtained from the linear fit of yield strength and strain rate values are plotted for both quasistatic ($m_{QSR}$) and HSR ($m_{HSR}$) regimes. The yield strength values of sx Ni show higher scatter compared to nc Ni. This is due to the more stochastic distribution of pre-existing dislocations in the sx Ni micropillar sample. The strain rate sensitivity exponent $m_{QSR}$ of sx Ni in the quasistatic regime ($10^{-3}$ to 1 $s^{-1}$) was determined to be 0.001, which is in line with the *m* values from SRJ tests reported in [20]. A noticeably higher $m_{HSR}$ value of 0.03 was observed in the HSR regime (1 to $10^2$ $s^{-1}$) suggesting higher dislocation-dislocation interactions and possible substructure formation. This is in line with increased strain hardening observed from the stress-strain curves in Fig. 1a earlier. The generated dislocations have less time to escape to the pillar free surface and get entangled, increasing both strain hardening behavior and strain rate sensitivity exponent at high strain rates (>1 $s^{-1}$). This trend also matches with the observations of Zhao et al. [29] who reported increase in dislocation density with strain rate of testing in copper micropillars, albeit over two orders of strain rates. They also reported extensive dislocation cell structure formation with cell sizes of ~ 1μm in Cu pillars with diameters of 3 and 5um. The overall *m* value for sx Ni over the tested strain rate range, from $10^{-3}$ to $10^2$ $s^{-1}$, was observed to be 0.011, which is ~ 10 times higher than $m_{QSR}$ value. The apparent activation volume ($V_{app}$) for sx Ni was found to be ~ 202$b^3$ which suggests extensive dislocation glide and forest hardening mechanism responsible for deformation. Fig 2a shows the post deformed images of the sx Ni micropillar at both quasistatic and HSR regimes. The sx Ni shows the characteristic slip behavior with distinct slip steps confined to the top 50% of the pillar height at both strain rate regimes. The prominent slip plane was identified to be of the {1 1 1} ⟨1 1 0⟩ type on the surface of the pillar. No noticeable differences were observed in the slip patterns for sx Ni across the tested strain rates, as seen Fig. 2a. While this suggests that the fundamental operative deformation mechanism of dislocation glide does not change at higher strain rates, the increased forest hardening and dislocation densities do end up altering the strain rate sensitivity exponent.

The nc Ni sample shows pronounced strain rate sensitivity in the quasistatic regime in comparison to sx Ni, due to smaller grain size. The *m* value of nc Ni in quasistatic regime was determined to be 0.018, which is one order of magnitude higher than that of sx Ni. We performed strain rate jump (SRJ) tests on additional nc Ni micropillars with strain rates varying between $10^{-3}$ $s^{-1}$ and $10^{-1}$ $s^{-1}$. These SRJ tests yielded *m* value of 0.013 (*supplementary information S3*) which is in good agreement with the $m_{QSR}$ observed from constant strain rate tests. For strain rates > 1 $s^{-1}$, we obtain $m_{HSR}$ of 0.008, which is much lower than $m_{QSR}$. This can be due to yield strength saturation to ~ 2.9 GPa beyond 1 $s^{-1}$. This yield strength saturation at high strain rates lowers the overall *m* value to 0.011. Interestingly, this trend is exactly opposite to what was observed in sx Ni micropillars. The overall *m* value obtained lies in the range of values previously reported for nanocrystalline Ni with similar grain size, although at low strain rates ($10^{-3}$ to $10^{-1}$ $s^{-1}$) [36]. Fig. 1c also shows micropillar compression data on nc Ni from Guillonneau et al. [6] for comparison. While the yield strength values are lower due to a different nc Ni sample used, the *m* value matches well with our results. They have also reported yield stress saturation for strain rates between 5 $s^{-1}$ and $10^2$ $s^{-1}$. It should be noted that both sx Ni and nc Ni show similar overall *m* value. The only explanation for this can be the increased contribution of dislocation substructure formation and forest hardening in sx Ni at high strain rates that raises its overall strain rate sensitivity. Bulk scale mechanical tests on nc Ni typically show a constant strain rate sensitivity up to 10 s$^{-1}$ and a sharp increase thereafter [37–39]. In contrast, we observe a decrease in strain rate sensitivity from 1s$^{-1}$ to 10$^3$ s$^{-1}$ due to strain localization in small pillar volumes.

The post deformed nc Ni micropillar images, shown in Fig. 2b, reveal subtle changes in surface features. While it was not possible to identify slip steps on the pillar surface due to extremely small grain size of ~30 nm, prominent grain boundary sliding was observed. The pillars show plastic deformation in the top half, consistent with the small taper in the pillars due to FIB milling. The deformation in the micropillars in the quasistatic strain rate regime, from 10$^{-3}$ s$^{-1}$ to 1 s$^{-1}$, appears to be more homogenous, with extensive grain boundary sliding occurring in the top half of the pillar. However, with further increase in strain rates, from 1 s$^{-1}$ to 10$^3$ s$^{-1}$, the deformation becomes more localized near the center of the micropillar. This strain localization seems to be the primary reason for yield strength saturation to ~ 2.9 GPa and for the decrease in strain rate sensitivity exponent at high strain rates. Previous studies suggest extensive dislocation mediated plasticity inside the grains with continuous nucleation and annihilation of dislocations at grain boundaries, which are assisted by grain boundary diffusion [33]. At quasistatic strain rates, there is sufficient time for thermally activated diffusion of atoms along grain boundaries to assist dislocation activity within the grains. On the pillar surface, this is often seen to result in grain boundary sliding and/or grain rotation. This explanation is in line with ~ 15b$^3$ activation volume observed from our micropillar tests and also matches with literature reported data [20,40,41]. However, at high strain rates, enough time is not available for grain boundary diffusion processes to assist the intra-granular dislocation plasticity throughout the pillar volume. Therefore, only limited sites can get activated resulting in strain localization, as observed in Fig. 2b. Strain localization can also occur due to localized adiabatic heating, which can lead to saturation of yield stress. nc Ni typically shows an increase in strain rate sensitivity exponent with temperature. Wang et al. [42] have reported an increase in the *m* value from 0.002 to 0.035 during a temperature rise from 296 K to 363 K respectively for uniform heating conditions. Since we observe a decrease in *m* value at high strain rates, we can clearly rule out significant adiabatic heating in the entire pillar volume at the highest tested strain rates. However, the possibility of extremely localized adiabatic heating at the grain boundaries always remains which can cause strain localization and this will be investigated in the next section through CP-FEM. Rather than a fundamental change in the operative deformation mechanism(s), strain localization seems to be primarily responsible for the observed decrease in strain rate sensitivity exponent at high strain rates. Nanotwinning, grain growth and extensive dislocation entanglements have been previously reported to dominate the deformation behavior of nc Ni at strain rates of 10$^6$ s$^{-1}$ [43], a regime that we clearly do not reach here.

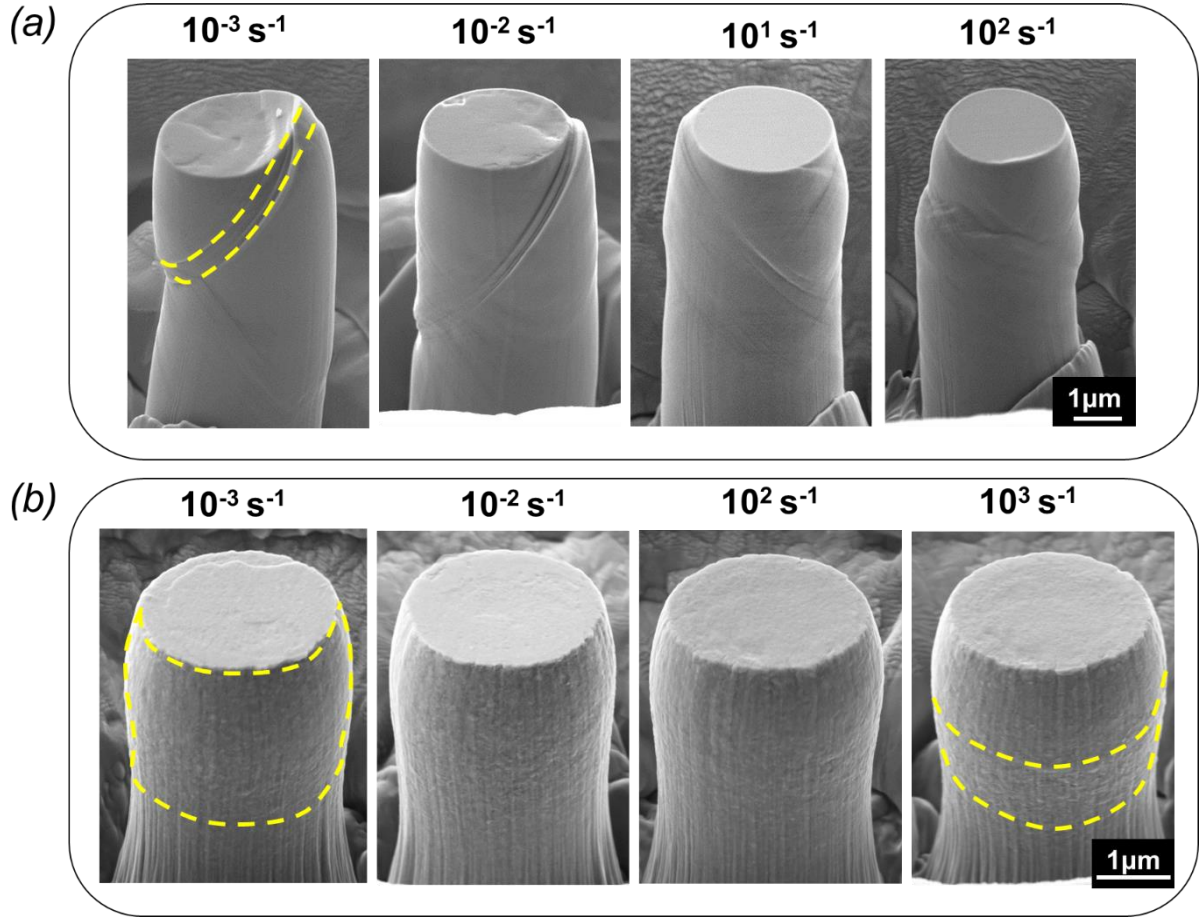

*Fig 2: Post deformed micropillars of a) sx nickel, b) nc nickel at quasistatic and HSR strain rate regimes, compressed to 20% nominal strains showing the deformed areas (marked in yellow). (Note: Post compressed images of micropillars for all strain rates tested are added in supplementary information S4 and S5)*

### 3.2 Adiabatic heat evolution at high strain rates

It can be seen from Fig 3a that the CP-FE model has a reasonable agreement with the experimental strain rate dependent yield and strain hardening behavior of the sx Ni. It is observed that in the HSR case, the model predicts very similar stress-strain behavior up to large strains irrespective of the taper angle of the micropillar. Fig 3b shows the experimental and simulated stress-strain curves for strain rates of $10^2$ s$^{-1}$ and $10^3$ s$^{-1}$ for nc Ni with high yield stress due to activated Hall-Petch effect in the model. Good agreement between the experiments and simulations are observed. Two different values of Taylor-Quiney coefficients were used to evaluate low and high heat generation at $10^3$ s$^{-1}$, corresponding to values 0.6 and 0.9 for variable $β$ [14]. Effectively, lower $β$ results in higher strain hardening since the CP model response is less affected by the limited temperature rise. Taylor-Quinney coefficient estimates from bulk scale tests are used to predict the extent of adiabatic heating in high strain rate microcompression. Since the applications of high strain rate testing at microscale is rapidly advancing, it is now very much necessary to understand the influence of the adiabatic temperature rise on the deformation behavior and any possible influence on the microstructural features. The effect of the microstructural features like grain boundaries on the energy storage was studied by Rittel et al. from high strain rate experiments [11]. The thermo-mechanical response of single crystal and polycrystalline Cu at high strain rates estimates that polycrystalline Cu stored more cold work than single crystal which underline the contribution of grain boundaries in storing the energy. Various continuum-based models were used to estimate the amount of energy stored and dissipated in the material and analyzed the effects of strain hardening and grain boundary sliding on the Taylor-Quiney coefficient [44,45]. The value of Taylor-Quiney coefficient depends on how the microstructural features like dislocation and grain boundaries corresponds to energy storage. Nieto-Fuentes et al. showed that the value of Taylor-Quiney coefficient is more influenced by the evolution of microstructure and the kinetic of dislocations [46]. This can be one of the reasons for increased strain hardening at high rates. However, it is difficult to track dislocation evolution and measurement of corresponding work to heat conversion using experiments. Using discrete dislocation plasticity framework, Benzerga et al. predicted the elastic energy locked in dislocation structures in in the crystal and estimated the heat dissipation during deformation [47]. With the help of molecular dynamics simulations, the dislocation interactions with dislocation and with grain boundaries can be tracked. Kositski et al. suggest that the grain boundaries play a significant role in the energy storage and release [48]. They employed molecular dynamics simulations to study the dislocation mechanisms by which the work is converted to heat.

CP-FE modeling provides extremely high spatial resolution in studying trends in adiabatic heating from nanometer/micrometer sized regions and along the grain boundaries, compared to what is currently not possible with thermal cameras. During micropillar compression of sx Ni in the quasistatic regime, it was observed that the apparent strain hardening of the material decreases rapidly, which can be caused by slip localization in the pillar. However, our simulations do not show slip localization in the pillar, which can be due to the assumption of identical initial dislocation density for all slip systems and their homogeneously distribution inside the pillar. Furthermore, the geometry of the pillar does not contain any irregularities, idealizing the micropillar compression test. Therefore, slip-slip interaction and length-scale regularization by the model does not predict nucleation of slip bands due to the absence of any slip nucleation sites. Instead, the simulation shows rather ideally distributed contribution from different slip systems for the [100] orientation. Fig 3c shows that the maximum predicted temperature increases inside the sx Ni micropillar is ~ 20 K when assuming fully adiabatic conditions for the simulation with Taylor-Quinney coefficient of 0.9. It also shows axial localization of dislocation density inside the pillar. This result suggests that the increase in temperature due to adiabatic heating is not significant, when notable slip band formation is not observed. Temperature rise can become higher if significant slip localization would occur in the simulations.

Fig. 3e shows the representative volume of polycrystalline Ni microstructure consisting of approximately 200 grains with grain size of ~30 nm that was used to predict the extent of adiabatic heating in nc Ni. The temperature distribution inside the polycrystalline microstructure of Ni at strain rate of $10^3$ s$^{-1}$ and $\beta = 0.9$ is shown in Fig. 3f (legend limited to 200 K increase). Significant temperature rise ranging from 30 to 170 K is predicted, whereas the peak temperature value exceeds 200 K locally at some hotspots near grain boundaries, at 15% axial strain (see *supplementary information S6* for the distribution of temperature change inside grains). However, the model does not account for the extensive grain boundary sliding observed in the experiments, which would have further reduced the temperatures at these grain boundary hotspots. These hotspots also show very high dislocation density in the deformed state (Fig. 3g) due to local temperature induced softening in the material behavior. Temperature dependent annihilation of dislocations is not sufficient to suppress the multiplication process with the identified model parameters. Generally, however, the temperature increase in the material can be judged to be ~ 70-150 K in majority of the grains, excluding grain boundary hotspots. Therefore, the simulations indicate that it is plausible that the temperature increase in polycrystalline micropillar tests can become very significant locally at strain rates of $10^3$ s$^{-1}$. It should be noted that these simulations provide the upper bound of the adiabatic temperature rise inside the micropillars, in the absence of heat conduction from the pillar to the indenter tip and base material, which are not considered here.

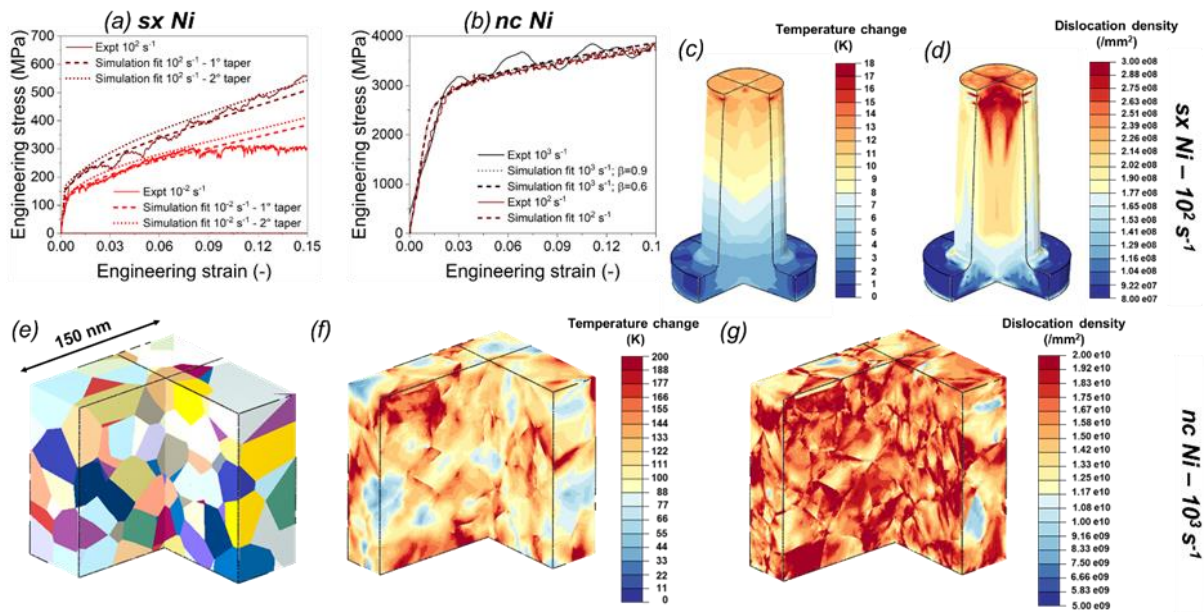

*Figure 3: a) Stress-strain comparison of simulations and experiments up to 15% nominal strain in a) sx Ni at the strain rates of $10^{-2}$ s$^{-1}$ and $10^2$ s$^{-1}$, b) nc Ni at the strain rate of $10^3$ s$^{-1}$. c) Distribution of temperature increase and d) dislocation density in sx Ni at $10^2$ s$^{-1}$ compression to 15% nominal strain for β = 0.9. e) Representative volume of the used polycrystalline microstructure of nc Ni. f) Distribution of temperature increase and g) dislocation density in nc Ni at $10^3$ s$^{-1}$ compression to 15% nominal strain for β = 0.9.*

The combination of adiabatic heating at the grain boundaries in nc Ni, predicted to rise up to 200 K in our CP-FEM calculations, and the mechanical work at the highest tested strain rate can potentially alter the microstructure resulting in grain growth. Grain growth has been observed in nc Ni after annealing for 1 h for temperatures as low as 200 ºC [49]. It should be noted that grain growth in nanocrystalline metals has been found to be highly dependent on the grain size and impurity content. Therefore, it becomes necessary to experimentally investigate any possible microstructural changes in our nc Ni. Fig. 4a shows the TKD analysis of a nc Ni pillar tested at $10^3$ s$^{-1}$. The inverse pole figure (IPF) map from a relatively undeformed region obtained from the bottom of the pillar shows roughly equiaxed grains. The IPF map of a deformed region obtained from the top of the pillar shows flattened grains due to compression. Note that large areas are unindexed due to extremely small grains and potentially large strains. The grain size distribution of the undeformed and deformed regions from the micropillar is shown in Fig. 4b and does not show any significant difference. There is barely any change in the average grain size ($\delta$) between the deformed and undeformed regions suggesting no grain growth in nc Ni tested at $10^3$ s$^{-1}$ due to adiabatic heating. This suggests that the extent of localized adiabatic heating at the grain boundaries is insufficient to drive substantial microstructural changes. Wang et al. also observed no breakdown in the nanocrystalline grains from ultrahigh strain rate testing at $10^7$ s$^{-1}$ performed on nc Ni and with predicted temperature increase of ~ 300 K [50]. The issue of adiabatic heating is important at high strain rate tests from the implication of microstructural changes occurring in the tested material. However, in such a short time period of deformation, which is of the order of few microseconds in our case, it seems that it is not enough to drive grain growth in nc Ni during the deformation.

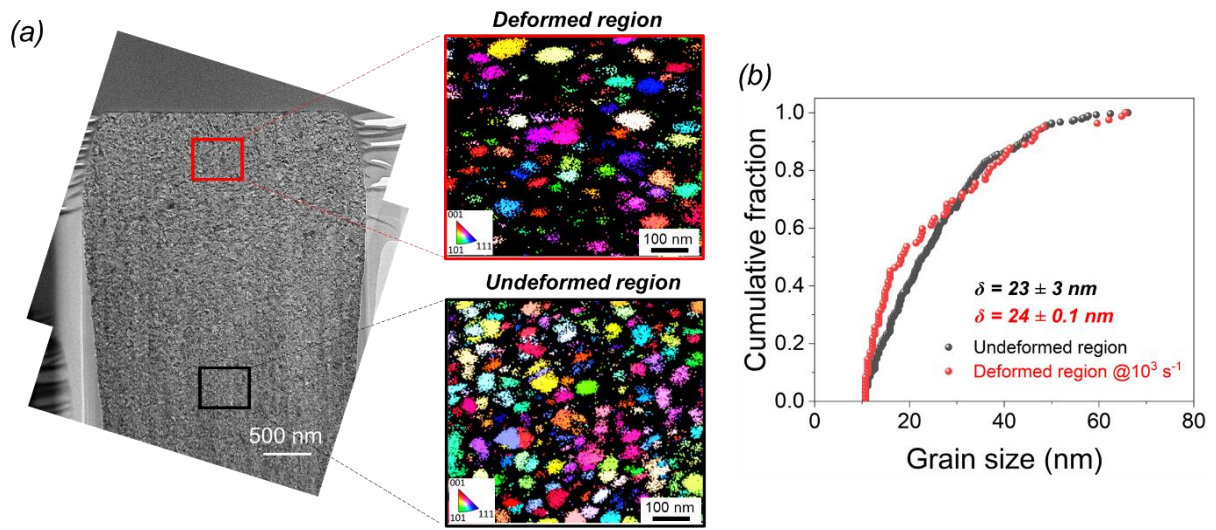

*Fig 4: a) Overview of deformed nc Ni micropillar with the IPF maps in the undeformed and deformed regions (Note: Locations of TKD scans are marked in the squares). b) Grain size distribution in the undeformed and deformed region of nc Ni from TKD analysis.*

## 4 Conclusions

This study reports the deformation response of sx and nc Ni over seven orders of strain rates ($10^{-3}$ s$^{-1}$ to $10^3$ s$^{-1}$) using micropillar compression experiments. sx Ni pillars predominantly deformed through dislocation glide whereas the deformation of nc Ni pillars showed grain boundary assisted dislocation plasticity with extensive grain boundary sliding observed on the surface. The overall strain rate sensitivity exponents were estimated to be 0.011 for both sx Ni and nc Ni. The higher-than-expected SRS of sx Ni appears to be driven by enhanced dislocation-dislocation interactions and possibly, substructure formation inside the pillar at high strain rates resulting in high m$_{HSR}$. nc Ni showed homogenous deformation in the top half of the pillar at quasistatic strains rates. At high strain rates (> 1s$^{-1}$), nc Ni pillars showed more inhomogeneous and localized plasticity resulting in yield stress saturation and lower m$_{HSR}$. In summary, sx Ni showed enhanced strain rate sensitivity while nc Ni showed decreased strain rate sensitivity at high strain rates (>1 s$^{-1}$). Crystal plasticity-based FEM simulations indicate temperature rise of only ~ 20 K in single crystal micropillars deformed at $10^2$ s$^{-1}$. However, nanocrystalline Ni pillars showed temperature rise of ~ 200 K locally near grain boundaries at $10^3$ s$^{-1}$, assuming perfectly adiabatic conditions. TKD analysis of a nc Ni pillar compressed at the highest tested strain rate ($10^3$ s$^{-1}$) did not show any noticeable grain growth suggesting that changes in strain rate sensitivity beyond 1 s$^{-1}$ is primarily due to strain localization. This is the first study to apply comprehensive CP-FE modeling to high strain rate micropillar compression of metals to investigate adiabatic heating related temperature rise at high strain rates. This study also suggests that adiabatic heating in micropillars should not be a major cause of concern for metals having moderate to high melting points.

## CRediT authorship contribution statement

**Nidhin George Mathews:** Conceptualization, Data curation, Formal Analysis, Investigation, Methodology, Visualization, Writing – original draft, Writing – review & editing. **Matti Lindroos:** Formal Analysis, Investigation, Writing – original draft, Writing – review & editing. **Johann Michler:** Writing – review & editing. **Gaurav Mohanty:** Conceptualization, Supervision, Funding acquisition, Writing – original draft, Writing – review & editing.

## Declaration of Competing Interest

The authors declare that they have no known competing financial interests or personal relationships that could have appeared to influence the work reported in this paper.

## Acknowledgements

N.G.M. and G.M. acknowledge financial support from Research Council of Finland (RCoF grant no. 341050). This work made use of H2MIRI (funded through RCoF grant number 353235) and Tampere Microscopy Center facilities at Tampere University.

# Supplementary information

## Deformation and adiabatic heating of single crystalline and nanocrystalline Ni micropillars at high strain rates

**Nidhin George Mathews[1,a], Matti Lindroos[2], Johann Michler[3], Gaurav Mohanty[1]**

[1] *Materials Science and Environmental Engineering, Tampere University, Korkeakoulunkatu 6, Tampere 33014, Finland.*

[2] *VTT Technical Research Centre of Finland Ltd, Tekniikantie 21, Espoo 02044, Finland.*

[3] *Laboratory of Mechanics of Materials and Nanostructures, Empa, Swiss Federal Laboratories for Materials Science and Technology, Feuerwerkerstrasse 39, Thun 3602, Switzerland.*

[a] *Corresponding author: nidhin.mathews@tuni.fi*

## S1. Transmission electron microscope characterisation

Transmission electrom microscope (TEM) characterisation was perfomed on the undeformed region of the nanocrystalline nickel sample inorder to determine its grain strucutre and nanocrystallinity. Fig. S1 shows the bright field TEM micrographs of the nanocrystalline grains along with the corresponding selected area diffraction (SAED) pattern from the same region. The SAED shows continuous ring pattern confirming the nanocrystallinity and random orientation of grains in the sample.

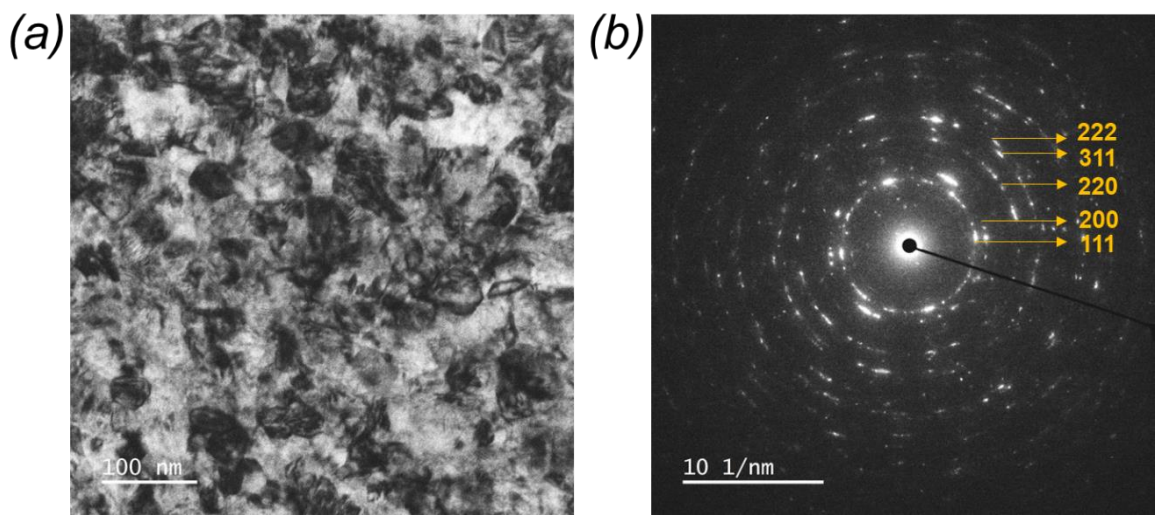

*Fig S1: a) Bright field TEM image showing the nanocrystalline grain structure and b) selected area electron diffraction pattern of nc Ni sample.*

## S2. Crystal plasticity modeling

*Table S1: Crystal plasticity parameters used in the simulations.*

| Parameters | Value | Unit |
|---|---|---|
| Elastic constant ($C_{11}$) | $248.33 \times 10^3 - 51.07 \cdot T$ | MPa |
| Elastic constant ($C_{12}$) | $151.2 \times 10^3 - 12.77 \cdot T$ | MPa |
| Elastic constant ($C_{44}$) | $125.7 \times 10^3 - 44.39 \cdot T$ | MPa |
| Shear modulus ($\mu$) | $\sqrt{C_{44}\left(\dfrac{C_{11}-C_{12}}{2}\right)}$ | MPa |
| Viscous parameter ($K$) | 18.0 | $MPa \cdot s^{1/N}$ |
| Strain rate exponent ($N$) | 12.0 | - |
| Hall-Petch coefficient ($K_{HP}$) | 3.7944 (polycrystal), 0.0 (micropillar) | $MPa/\sqrt{mm}$ |
| Grain size ($d$) | 30 nm (polycrystal), and the average diameter (micropillar) | μm |
| Length of Burgers vector ($b^s$) | 0.254 | nm |
| Interaction coefficients ($a_i$) | $a_1 = 0.124, a_2 = 0.124, a_3 = 0.07, a_4 = 0.625, a_5 = 0.137, a_6 = 0.122$ | - |
| Annihilation distance (reference) ($y_0$) | $5 \cdot b^s$ | nm |
| Capture radius energy ($A_{rec}$) | $2 \times 10^{-20}$ | J |
| Reference slip rate ($\dot{\gamma}_0$) | $1 \times 10^7$ | $s^{-1}$ |
| Forest hardening parameter | 70 | - |
| Coplanar hardening parameter | 150 | - |
| Length-scale penalty modulus ($H_\chi$) | 2000 | $MPa \cdot mm^2$ |
| Length-scale parameter ($A$) | 0.008 | N |
| Mass density ($\rho$) | 8830 | $kg/m^3$ |
| Heat capacity ($c$) | 455 | J/K |

## S3. Strain rate jump tests on nanocrysalline nickel

Strain rate jump tests on nanocrystalline nickel micropillars were performed by varying the strain rate by one order of magnitude for each jump. Fig. S2 shows representative engineeering stress-strain curve for strain jumps between $10^{-1}$ s$^{-1}$ to $10^{-3}$ s$^{-1}$. The strain rate sensitivity (*m*) value of 0.013 was estimated using Eq. 10 (in the main manuscript).

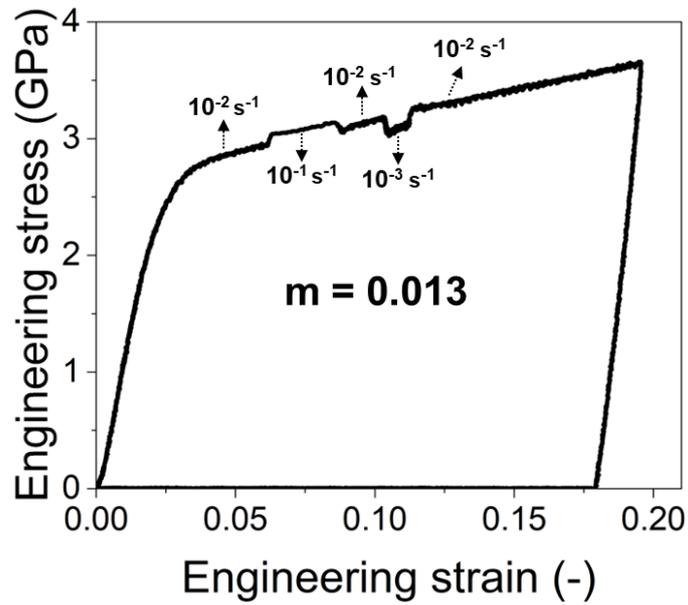

*Fig S2: Representative engineering stress-strain curve of strain rate jump tests on nc Ni micropillars.*

**S4. Micropillar compression of sx Ni at different orders of strain rates**

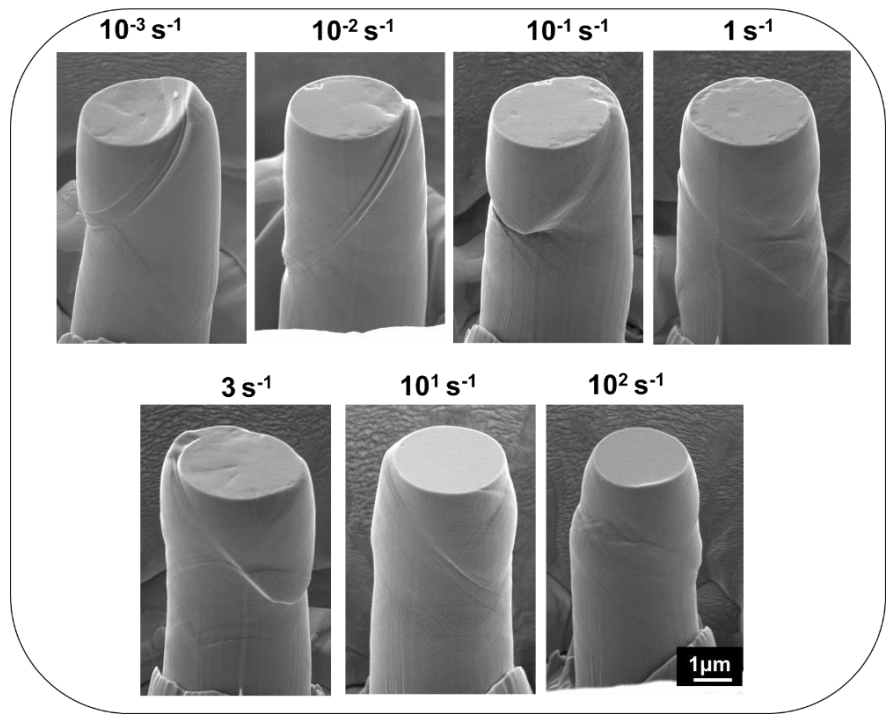

*Fig S3: Post-compression images of sx Ni micropillars ranging from $10^{-3}$ s$^{-1}$ to $10^{2}$ s$^{-1}$.*

## S5. Micropillar compression of nc Ni at different orders of strain rates

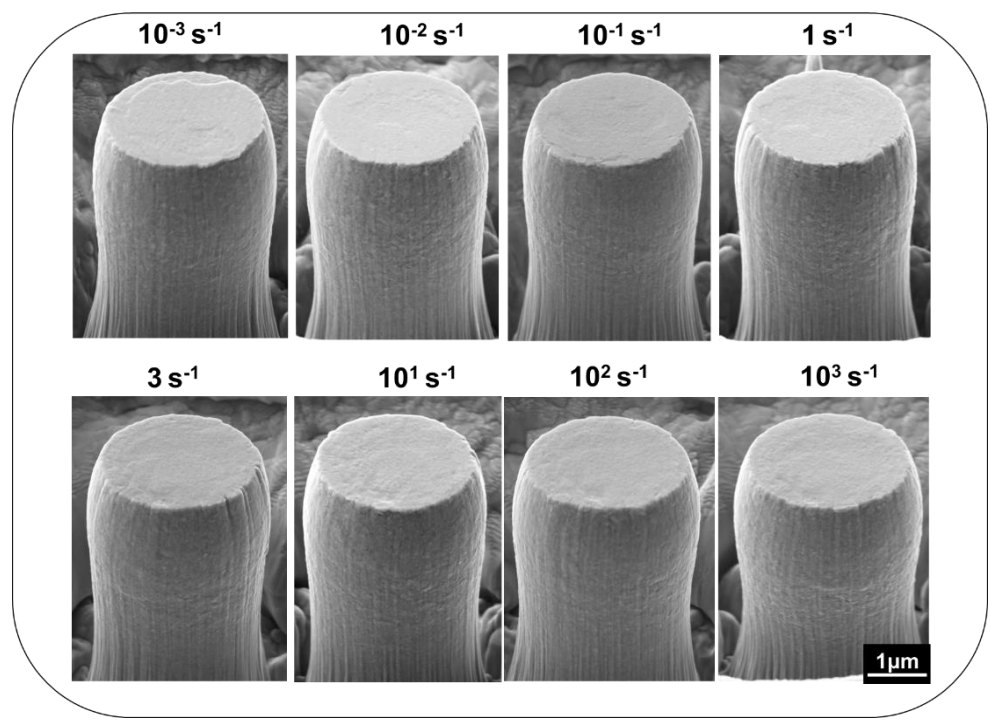

*Fig S4: Post-compression images of nc Ni micropillar ranging from $10^{-3}$ $s^{-1}$ to $10^3$ $s^{-1}$.*

## S6. Temperature change inside the nc Ni grains from CP-FEM simulations

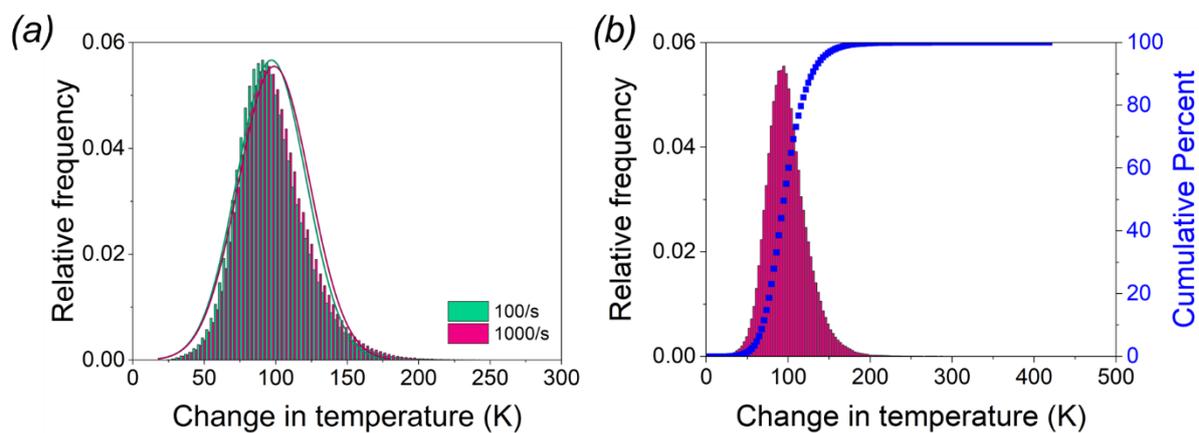

*Fig S5: a) Temperature rise distribution inside the nc Ni grain estimated from CP-FEM simulations at $10^2$ $s^{-1}$ and $10^3$ $s^{-1}$ strain rates at 15% strain. b) Cumulative percent of the temperature rise in nc Ni at $10^3$ $s^{-1}$ strain rates at 15% strain.*